# Antipolar transitions in GaNb$_4$Se$_8$ and GaTa$_4$Se$_8$

M. Winkler[1], L. Prodan[1], V. Tsurkan[1,2], P. Lunkenheimer[1,*], and I. Kézsmárki[1]

[1]*Experimental Physics V, Center for Electronic Correlations and Magnetism, University of Augsburg, 86159 Augsburg, Germany*
[2]*Institute of Applied Physics, Chisinau MD-2028, Republic of Moldova*

We present dielectric, polarization, resistivity, specific heat, and magnetic susceptibility data on single crystals of the lacunar spinels GaNb$_4$Se$_8$ and GaTa$_4$Se$_8$, tetrahedral cluster-based materials with substantial spin-orbit coupling. We concentrate on the possible occurrence of antipolar order in these compounds, as previously reported for the isoelectronic GaNb$_4$S$_8$, where spin-orbit coupling plays a less important role. Our broadband dielectric-spectroscopy investigations reveal clear anomalies of the intrinsic dielectric constant at the magneto-structural transitions in both systems that are in accord with the expectations for antipolar transitions. A similar anomaly is also observed at the cubic-cubic transition of the Nb compound leading to an intermediate phase. Similar to other polar and antipolar lacunar spinels, we find indications for dipolar relaxation dynamics at low temperatures. Polarization measurements on GaNb$_4$Se$_8$ reveal weak ferroelectric ordering below the magneto-structural transition, either superimposed to antipolar order or emerging at structural domain walls. The temperature-dependent dc resistivity evidences essentially thermally-activated charge transport with different activation energies in the different phases. A huge step-like increase of the resistivity at the magneto-structural transition of the Ta compound points to a fundamental change in the electronic structure or the mechanism of the charge transport. At low temperatures, charge transport is governed by in-gap impurity states, as also invoked to explain the resistive switching in these compounds.

## I. INTRODUCTION

In recent years, materials where non-canonical mechanisms generate polar or antipolar order have been attracting considerable interest [1,2,3,4,5,6,7,8,9]. For example, such systems often tend to exhibit simultaneous polar and magnetic order, i.e., they are multiferroic with high potential for future information technologies. Lacunar spinels with the general formula $AM_4X_8$ are prominent examples of such systems [5,10,11,12,13,14]. Interestingly, their crystalline lattice comprises cubane-type $M_4X_4$ clusters that can be regarded as molecular units with specific spin and orbital degrees of freedom [15,16,17,18,19,20]. In several members of this material class, including GaV$_4$S$_8$, GaV$_4$Se$_8$, GaMo$_4$S$_8$, and GeV$_4$S$_8$, ferro-type polar order was found [10,11,12,14,21,22,23]. It is believed to be driven by orbital ordering below the Jahn-Teller (JT) transition and leads to a structural distortion of the clusters, generating an electric dipolar moment. Since the distortion direction is the same for all clusters in these materials below the JT transition [15], polar order arises. For some of these systems, switchability of the polarization was demonstrated, making them JT-effect driven ferroelectrics [10,11,12,22]. Moreover, at low temperatures they exhibit various magnetic orderings, including Néel-type skyrmion lattice states [10,11,24,25,26,27,28,29,30], thus becoming multiferroic with significant magnetoelectric effects.

In contrast, in the lacunar spinel GaNb$_4$S$_8$, very recently strong hints at antipolar order were found and it was proposed to represent the first example of cluster JT effect-driven antiferroelectricity [31]. Below the JT transition of this system, the individual tetrahedral Nb$_4$ clusters undergo a polar distortion [32], similar to the distortion of the V$_4$ clusters in GaV$_4$S$_8$ and GaV$_4$Se$_8$ [15,33]. However, in marked contrast to these ferroelectric lacunar spinels, the distortion of the Nb$_4$ clusters shows a staggered pattern, i.e. the polar distortion varies between neighboring clusters [32] (see, e.g., Fig. 1(d) in Ref. [31]). The concomitant structural transition of GaNb$_4$S$_8$ from space group $F\bar{4}3m$ to $P2_12_12_1$ is consistent with the onset of antiferroelectricity [31,34]. This material also differs from its sister compounds with ferro-type polar order by revealing a non-magnetic ground state due to spin-singlet formation below the structural transition [35].

Two materials that are closely related and isoelectronic to GaNb$_4$S$_8$, namely GaNb$_4$Se$_8$ and GaTa$_4$Se$_8$ [36], have recently attracted considerable interest due to an electric-field induced Mott transition and the occurrence of superconductivity under pressure [37,38,39,40]. Their clusters are known to also distort along different directions below structural transitions as schematically indicated in the lower inset of Fig. 1(a) [41]. The lowest-temperature space group of GaNb$_4$Se$_8$ is $P2_12_12_1$ [41], just as for GaNb$_4$S$_8$. For GaTa$_4$Se$_8$, $P\bar{4}m2$ [41] and $P\bar{4}2_1m$ [42,43] low-temperature structures were considered. However, recent single-crystal x-ray diffraction measurements of this compound also point to $P2_12_12_1$ [14,44]. In both materials, below their magneto-structural transition, taking place at $T_m \approx 33$ K (Nb) and 53 K (Ta), a non-magnetic spin-singlet ground state emerges, again similar to GaNb$_4$S$_8$ [41]. It was noted to be consistent with the formation of a valence bond solid state [41]. Remarkably, in GaNb$_4$Se$_8$ an intermediate phase with space group $P2_13$ was found to arise between

*Corresponding author: peter.lunkenheimer@physik.uni-augsburg.de

$T_Q \approx 50$ K and $T_m$ [41]. It was ascribed to the appearance of "electric quadrupolar order" [41]. It should be noted that the cooperative staggered distortions of the $Nb_4$ clusters, as depicted in the lower inset of Fig. 1(a), are expected to appear already at $T < T_Q$ [41], though the symmetry is further lowered below $T_m$. Interestingly, the space group $P2_13$ is also found in the skyrmion-lattice host $Cu_2OSeO_3$ [45], and, based on dielectric and polarization measurements, it was speculated that "complex antiferroelectric order" may form in this system [46].

For $GaNb_4S_8$ the JT effect seems to be the dominant driving force of its magneto-structural transition [31]. In contrast, based on magnetic-susceptibility investigations of polycrystalline samples, Ishikawa et al. [41] recently claimed that in $GaNb_4Se_8$ and $GaTa_4Se_8$ spin-orbit coupling also plays an important role and it could be the realization of a "molecular $j_{eff}$ state" [47,48]. This confirmed earlier conclusions for the Ta compound by Kawamoto et al., also drawn from magnetic-susceptibility measurements [49]. The significance of spin-orbit coupling was later also verified for single-crystalline samples of both compounds by Petersen et al. [50].

The main purpose of the present work is to check for the possible signatures of antiferroelectricity in single-crystalline $GaNb_4Se_8$ and $GaTa_4Se_8$ using dielectric spectroscopy and polarization measurements. To overcome the obstacle of dominating non-intrinsic contributions to the dielectric response of these essentially semiconducting samples [51,52], dielectric spectra were collected in a broad frequency range up to about 1 GHz. Moreover, we characterize the phase transitions in single-crystalline samples by specific-heat and magnetic-susceptibility measurements and provide detailed results on the dc resistivity in a broad temperature range.

## II. EXPERIMENTAL DETAILS

Polycrystalline samples of $GaNb_4Se_8$ and $GaTa_4Se_8$ were synthesized by solid-state reactions from the high-purity elements: Ga (99.9999%), Nb (99.8%), Ta (99.9%), and Se (99.999+%). The powder was sealed in an evacuated quartz ampoule and heated for 72 hours at 900 ºC. Two subsequent sintering steps were necessary to obtain the single ternary phase. The polycrystalline material was used as a source for single-crystal growth by chemical transport reactions. The growth was performed in closed quartz ampoules at temperatures between 1000 and 950 °C using iodine as transport agent. The high-quality single crystals having the shape of truncated octahedrons of size up to 3 mm with shiny (111) and (001) planes [see insets of Figs. 1(b) and 2(b) for examples] were obtained within two months of transport. For the electrical measurements, single crystals with well-defined (100) faces were polished down to arrive at a platelet-shaped geometry. For the dielectric and polarization experiments, contacts were applied to opposite sides of these crystals using silver paint, leading to a capacitor-like geometry. For the four-point dc measurements, coplanar, stripe-shaped contacts were applied.

The polarization measurements were performed in an Oxford Helium-flow cryostat using an electrometer (Keysight B2987A). First, electric poling fields were applied during cooling from 35 K down to 4 K. Before warming and measuring the pyroelectric current, the sample was short-circuited for at least one minute. The pyroelectric current was then recorded upon heating with a constant rate of ~ 7 K/min. The polarization was calculated by integration of the current with respect to time. The dielectric response in a frequency range between 1 Hz and several MHz was determined by a frequency-response analyzer (Novocontrol Alpha Analyzer). Additional high-frequency measurements were carried out in a range between 1 MHz and about 1 GHz using an I-V coaxial technique employing an impedance analyzer (Keysight E4991B) [53]. For cooling and heating, a $^4$He-bath cryostat (Cryovac) was used. The dc resistivity was measured in four-probe geometry at temperatures between about 70 and 300 K and in two-point configuration at lower temperatures. This was done in a physical properties measurement system (Quantum Design PPMS) in a temperature range of 4 to 200 K using the same devices as for the polarization measurements.

The heat capacity of single crystals taken from the same batch as the electric measurements was investigated upon heating using the PPMS at temperatures between 2 and 300 K in magnetic fields up to 9 T. Magnetic-susceptibility measurements were performed in a superconducting quantum interference device (SQUID) magnetometer (Quantum Design MPMS 3), where the external static magnetic field of 1 T was applied parallel to the crystallographic $\langle 100 \rangle$ and $\langle 111 \rangle$ directions.

## III. RESULTS AND DISCUSSION

To characterize the phase transitions in the investigated single crystals, we investigated the specific heat $C_p$ and magnetic susceptibility $\chi$ of $GaNb_4Se_8$ (Fig. 1) and $GaTa_4Se_8$ (Fig. 2) as a function of temperature. In $C_p(T)$, the known phase transitions of these compounds are well reproduced and lead to extremely sharp peaks as shown in more detail in the upper insets of Figs. 1 and 2. Their much narrower peak shape and higher amplitude, compared to the previously published results on polycrystals [41,49], evidences the high quality of our samples and points to the first-order nature of these transitions. The obtained transition temperatures are $T_Q \approx 48$ K and $T_m \approx 31$ K for $GaNb_4Se_8$ and $T_m \approx 52$ K for $GaTa_4Se_8$, in reasonable agreement with the results in Ref. [41]. As revealed by the upper insets of Figs. 1 and 2, there is no significant influence of an applied external magnetic field of 9 T on the phase transitions.

The temperature-dependent magnetic susceptibility data shown in Figs. 1(b) and 2(b) closely resemble those reported in Ref. [41] for both materials and point towards singlet formation below the magneto-structural transition. Very similar results measured in our group at a magnetic field of 7 T, can be found in Ref. [50]. A quantitative analysis of those data corroborates the mentioned conclusions by Ishikawa et al. [41] on the significant spin-orbit coupling in these systems. The upturn of $\chi(T)$ at the lowest temperatures can be ascribed to free spins from impurities or defects [41] (see Appendix A for more details).

Figure 3 shows the temperature-dependent dielectric constant $\varepsilon'$ (a) and the real part of the conductivity $\sigma'$ (b) as measured for GaNb$_4$Se$_8$ at various frequencies. Due to the relation $\sigma'/\nu \propto \varepsilon''$, Fig. 3(b) also provides information on the temperature dependence of the dielectric loss $\varepsilon''$. This material is semiconducting which can lead to the formation of Schottky diodes at the interfaces between the bulk sample and the applied metallic contacts. They can produce strong non-intrinsic contributions to the dielectric properties [51,52]. Indeed, the unrealistically large absolute values of the dielectric constant of the order of several thousand, detected at the higher temperatures and low frequencies in Fig. 3(a), clearly points to such effects. $\varepsilon'(T)$ in this region exhibits a smooth step-like decrease to lower values with decreasing temperature and the steps shift to lower temperatures for lower frequencies. This signifies a relaxational process. In the present case, it can be ascribed to a non-intrinsic, so-called Maxwell-Wagner (MW) relaxation, which arises when the insulating depletion layer of the Schottky diode is short-circuited by the associated capacitance at high frequencies or low temperatures (see, e.g., Refs. [51,52] for a detailed discussion). As mentioned in section II, the dielectric data were measured using two different experimental setups employed below and above about 1 MHz, which required different sample contacting. The values of the high-temperature plateau in Fig. 3(a), obtained from the two measurement runs, significantly differ, corroborating the non-intrinsic, contact-related character of the $\varepsilon'$ data in this region.

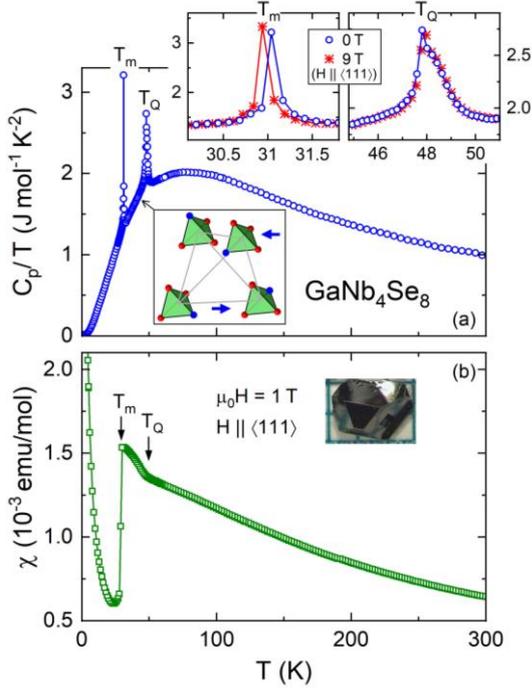
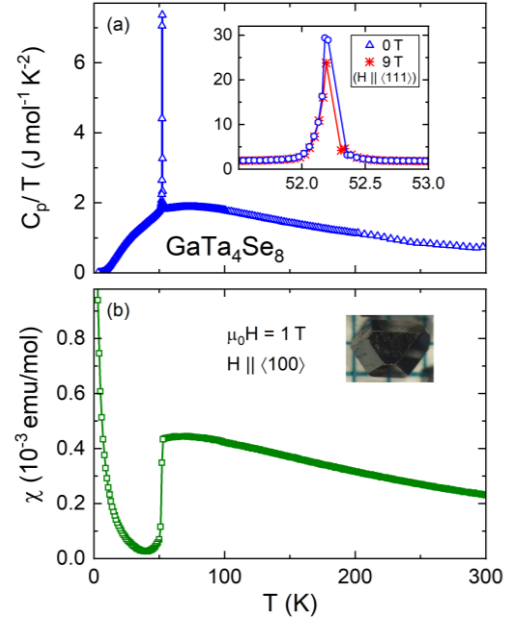

FIG. 1. (a) Temperature-dependent molar heat capacity of GaNb$_4$Se$_8$ plotted as $C_p/T$ vs $T$ for clarity reasons. The upper insets provide zoomed views of the phase-transition regions. The stars show additional data measured under magnetic fields of 9 T, applied along $\langle 111 \rangle$. For the intermediate $P2_13$ phase, the lower inset in (a) schematically depicts the arrangement of four adjacent Nb$_4$ clusters which are distorted along one tip of the tetrahedron, indicated in blue [41]. The distortions of the individual clusters point along different directions. The arrows indicate the approximate direction of the summed-up distortions for the upper and lower clusters, rationalizing antipolar order. (b) Temperature dependence of the magnetic susceptibility of GaNb$_4$Se$_8$ with the transition temperatures indicated by arrows. The inset shows a typical single crystal.

FIG. 2. (a) Temperature-dependent $C_p/T$ of GaTa$_4$Se$_8$. The upper inset provides a zoomed view of the phase-transition region. The stars show additional data measured under magnetic fields of 9 T, applied along $\langle 111 \rangle$. (b) Temperature dependence of the magnetic susceptibility of GaTa$_4$Se$_8$. The inset shows a typical single crystal.

As seen in Fig. 3(a), the two phase transitions in GaNb$_4$Se$_8$ lead to distinct anomalies in $\varepsilon'(T)$ at all frequencies. However, only for $\nu \geq 97$ MHz the MW-relaxation step has shifted sufficiently far towards high temperatures to reveal the completely unobstructed intrinsic $\varepsilon'(T)$ behavior at $T_m$ and $T_Q$. For 812 MHz, it is shown in the inset of Fig. 3(a). Interestingly, both the suggested electric quadrupolar transition at $T_Q$ and the magneto-transition at $T_m$ give rise to a significant downward step in $\varepsilon'(T)$. A similar $\varepsilon'$ step was found at the JT transition in GaNb$_4$S$_8$ and taken as hint at antipolar ordering [31], based on theoretical expectations for $\varepsilon'(T)$ at antiferroelectric phase transitions [34,54]. Antipolar order of GaNb$_4$Se$_8$ upon entering the intermediate phase is also well consistent with the structural distortions arising below $T_Q$ indicated in the lower inset of Fig. 1(a) [41]. The second $\varepsilon'(T)$ step observed at $T_m$ in GaNb$_4$Se$_8$ indicates a further change in the dipolar arrangement at this magneto-structural transition. A detailed structural investigation is currently under way to provide a deep insight into the low-temperature antipolar structure of GaNb$_4$Se$_8$ and GaTa$_4$Se$_8$.

The conductivity $\sigma'(T)$, shown in Fig. 3(b), in general tends to increase with increasing temperature and reveals significant frequency dispersion. To understand this behavior, it is instructive to compare it with the separately measured dc conductivity $\sigma_{dc}(T)$ shown by the dashed line. Above about 70 K, these data were obtained in four-probe geometry and thus should not be hampered by electrode contributions. For $T \gtrsim 50$ K, $\sigma'(T)$ at low frequencies is significantly smaller than $\sigma_{dc}(T)$, reflecting the fact that it is dominated by non-intrinsic contact contributions in this region. Only at high frequencies, $\sigma'(T)$ reasonably agrees with $\sigma_{dc}(T)$ [55] because here the contact capacitances become short-circuited [51,52]. The shoulder observed, e.g., for the 9.85 MHz curve around 100 K corresponds to a crossover from the electrode-dominated to intrinsic behavior upon cooling. It is directly associated with the MW relaxation steps in $\varepsilon'(T)$ discussed above. Overall, such behavior is typical for non-intrinsic electrode effects as, e.g., discussed in detail in Ref. [52].

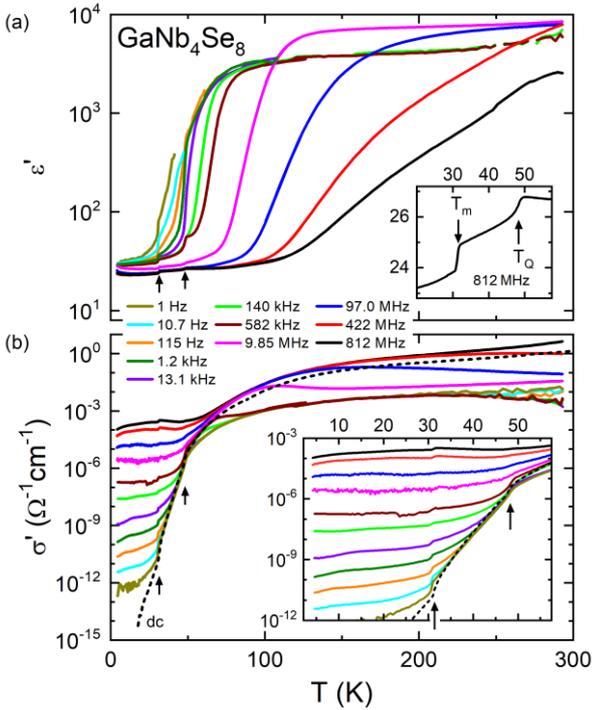

FIG. 3. Temperature-dependent dielectric constant (a) and conductivity (b) of GaNb$_4$Se$_8$, measured at various frequencies. The inset in (a) shows a magnified view of $\varepsilon'(T)$ at the highest frequency at temperatures around the phase transitions. The inset in (b) presents the low-temperature region of $\sigma'(T)$. The dashed line in (b) shows the separately measured dc conductivity. The arrows indicate the two phase transitions.

Below about 50 K, in general $\sigma'(T) \geq \sigma_{dc}(T)$ is found. Here the dc conductivity is sufficiently low to make non-intrinsic contributions negligible. Depending on frequency, with increasing temperature the different $\sigma'(T)$ curves merge with each other. This signifies a frequency-independent low-frequency region, which can be identified with the dc conductivity [52]. Indeed, it agrees with the directly measured $\sigma_{dc}(T)$ (dashed line). The dc conductivity, both from the dielectric and direct dc measurements, reveals two steplike reductions at the phase transitions by a factor of 2 - 3 [see inset of Fig. 3(b)]. The dc conductivity will be discussed in more detail below.

Interestingly, there is considerable *intrinsic* frequency dispersion in this low-temperature region with $\sigma'$ continuously increasing with $\nu$. The general trend is in accord with ac conductivity caused by hopping charge transport [56,57] as previously considered for GaNb$_4$Se$_8$, e.g., in Refs. [37,38]. Superimposed on this weakly temperature-dependent ac conductivity, partly faint indications of peaks or shoulders show up (e.g., at ~15 K for 13.1 kHz; see inset) that shift to higher temperatures with increasing frequencies. They indicate some intrinsic dipolar relaxation process in this region [58]. Relaxations below a dipole-ordering transition were also reported for the polar lacunar spinels GaV$_4$S$_8$ [59], GaV$_4$Se$_8$ [11], GaMo$_4$S$_8$ [60], GeV$_4$S$_8$ [61], and for the antipolar GaNb$_4$S$_8$ [14]. This was taken as evidence for the order-disorder character of their polar transitions. Unfortunately, in the present case the detected relaxation process is too weakly pronounced to allow for a meaningful quantitative evaluation. Notably, the magnetic phase transition leads to a jump in $\sigma'(T)$ at all frequencies, instead of showing up at low frequencies only, where dc transport dominates. Thus, the ac conductivity and/or dipolar dynamics are also affected by this transition.

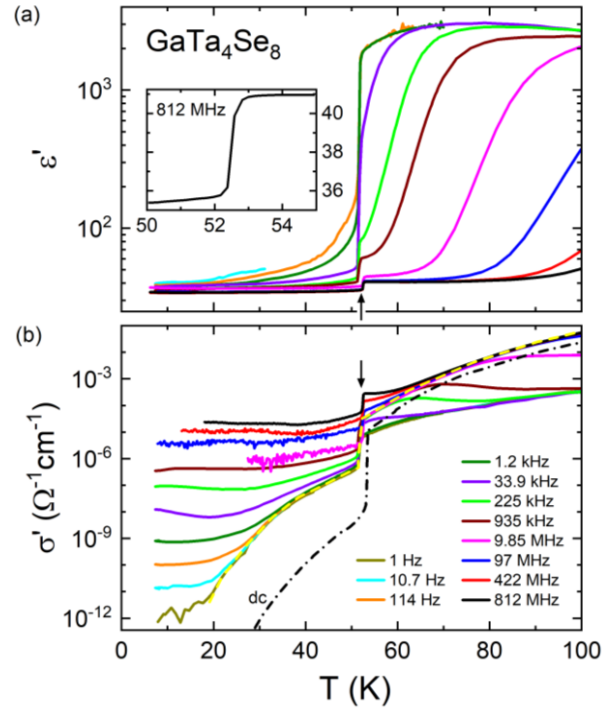

FIG. 4. Temperature-dependent dielectric constant (a) and conductivity (b) of GaTa$_4$Se$_8$, measured at various frequencies. The arrows indicate the phase transition. The inset in (a) shows a magnified view of $\varepsilon'(T)$ at the highest frequency at temperatures around the phase transition. The dash-dotted line in (b) shows the separately measured dc conductivity. The yellow dashed line indicates the dc conductivity as estimated from the dielectric data.

Figures 4(a) and (b) show $\varepsilon'(T)$ and $\sigma'(T)$ of GaTa$_4$Se$_8$, respectively (the plot is restricted to temperatures below 100 K; at $T > 100$ K a trivial continuation of the MW relaxation is observed). Just as for GaNb$_4$Se$_8$, at high temperatures and low frequencies non-intrinsic MW effects dominate the dielectric response. For all frequencies, the dielectric constant exhibits a strong anomaly at $T_m$. For $\nu \geq 97$ MHz, the frequency is sufficiently large to fully exclude any non-intrinsic contributions. As shown in the inset, the intrinsic $\varepsilon'(T)$ again reveals a clear step at the transition, suggesting an antipolar transition [34,54].

Above the transition, $\sigma'(T)$ of GaTa$_4$Se$_8$ [Fig. 4(b)] again displays the signatures of a MW relaxation. The yellow dashed line at $T > T_m$, into which the high-frequency $\sigma'(T)$ curves merge, indicates the dc conductivity estimated from these data [52]. At these high temperatures, it is in principle accord with the directly measured $\sigma_{dc}(T)$ but deviates by about a factor of 2. This deviation may well be due to the different contact geometries (electrodes on opposite sample faces vs coplanar four-point contacts) generating different field distributions [55]. For both data sets, a marked conductivity reduction appears below $T_m$ in this material, but this reduction is significantly stronger for the dc measurement. Consequently, at $T < T_m$ the directly measured $\sigma_{dc}$ is by about 2 - 3 decades smaller than the values estimated from the dielectric measurements. As the same sample was used for both low-temperature measurements, the origin of this discrepancy is unclear at present. Strong field-induced resistance switching is known to occur in this material for fields above about 1 - 2 kV/cm [40,62]. The significantly lower ac fields of about 80 V/cm applied during the dc measurements, preceding the dielectric ones (also using 80 V/cm), seems to exclude such switching. An influence of microcracks, appearing during cycling through the structural transition, also is unlikely because this should not *enhance* the low-temperature dc conductivity in the subsequent dielectric measurement run. Probably, the difference in the dc conductivity mirrors an anisotropy of $\sigma_{dc}$ appearing when the sample becomes non-cubic below $T_m$. Assuming a different domain distribution after crossing the transition in the two measurement runs then could explain the results. Irrespective of the dc conductivity, below the magneto-structural transition, $\sigma'(T)$ of GaTa$_4$Se$_8$ [Fig. 4(b)] exhibits qualitatively similar behavior as GaNb$_4$Se$_8$ [Fig. 3(b)]. Again, strong frequency dispersion and indications of peaks point to dipolar dynamics and hopping conductivity.

Figure 5 shows the temperature-dependent dc resistivity, $\rho_{dc} = 1/\sigma_{dc}$, of both investigated systems using an Arrhenius representation. In contrast to previously published data [38,39,40,63], the measured $\rho_{dc}(T)$ extends to temperatures below the phase transitions, where it starts to become too large to be detectable by the usual four-point devices. As mentioned in section II, thus there we used a two-point setup. As discussed above, for GaNb$_4$Se$_8$ the comparison of $\sigma_{dc}(T)$ with the $\sigma'(T,\nu)$ results in Fig. 3(b) clearly reveals that non-intrinsic contact contributions due to Schottky-diode formation can be neglected below about 50 K. For GaTa$_4$Se$_8$ [Fig. 4(b)], in principle the same can be said. The much lower directly measured $\sigma_{dc}$ compared to the dielectric results in this material [cf. dash-dotted and dashed lines in Fig. 4(b)] should rather diminish the likelihood of Schottky-diode formation.

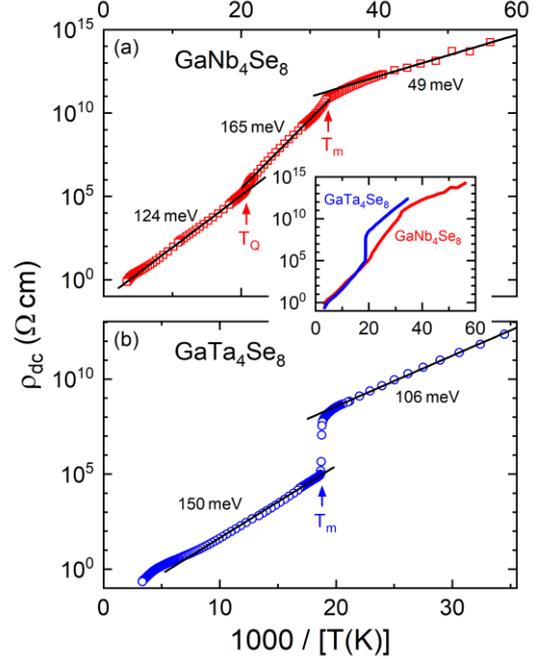

FIG. 5. (a) Arrhenius plot of the temperature-dependent dc resistivity of GaNb$_4$Se$_8$ (a) and GaTa$_4$Se$_8$ (b). The lines indicate fits with Eq. (1) yielding different activation energies as noted in the figure. To enable a direct comparison, the inset shows both $\rho_{dc}(T)$ curves within the same frame.

As revealed by Fig. 5, for both materials the phase transitions divide the $\rho_{dc}(T)$ traces into approximately linear regions with different slopes, implying thermally-activated charge transport with different activation energies $E_g$. Assuming intrinsic band conduction, the corresponding Arrhenius law is

$$\rho_{dc} = \rho_0 \exp[E_g/(2k_B T)] \qquad (1)$$

where $\rho_0$ is a pre-exponential factor and $k_B$ is the Boltzmann constant. The fits with Eq. (1) (lines in Fig. 5) reveal band-gap values between 49 and 165 meV as indicated in the figure. At both transitions of the Nb compound [Fig. 5(a)], the resistivity exhibits a steplike increase by about a factor of 2 - 3. The single transition in GaTa$_4$Se$_8$ [Fig. 5(b)] leads to a much stronger and abrupt rise of $\rho_{dc}$ by about three decades. Remarkably, this is not accompanied by an increase in the band gap, which even drops upon the transition. This points to a fundamental change of the charge-transport mechanism at $T_m$. Formally, this resistivity jump has to be ascribed to a variation of $\rho_0$ in Eq. (1) which depends, amongst others, on the number density of charge carriers. A possible explanation is a strong increase of the band gap due to the structural changes below $T_m$, making intrinsic band conductivity by charge-carrier excitation from the valence into the conduction band negligible. This would leave conductivity of charge carriers, excited from in-gap states arising, e.g., from slight impurities that act like dopants, as the dominant mechanism. The small number of such impurity electrons (or holes) then explains the increased resistivity, although their activation energy (from the in-gap impurity states to the conductivity band) is smaller. Band-structure calculations for the low-temperature phase of GaTa$_4$Se$_8$ would be necessary to check the validity of this scenario. An alternative explanation is the emergence of domain walls with small charge-carrier density below the structural transition, which have to be crossed by the charge carriers and thus dominate their resistivity at $T < T_m$.

For GaNb$_4$Se$_8$ [Fig. 5(a)], the resistivity increase below the cubic-cubic transition at $T_Q$ is consistent with an accompanying rise of $E$. However, below $T_m$ the activation energy again decreases. Therefore, for this transition a crossover to conduction arising from in-gap impurity states can be assumed, too. The inset of Fig. 5 demonstrates that, at high temperatures, both materials reveal quite similar $\rho_{dc}(T)$.

In literature, there are several reports on energy gaps deduced from resistivity measurements of GaNb$_4$Se$_8$ and GaTa$_4$Se$_8$ at temperatures above the phase transition(s) [37,38,39,40,63]. The results vary between 140 and 240 meV for the Ta and between 100 and 280 meV for the Nb compound [64]. However, part of these investigations are restricted to rather small temperature ranges around room temperature [38,40,63] where our data exhibit some deviations from Arrhenius behavior (Fig. 5). The values of 140 meV (Ta) [37,39,64] and 100 eV (Nb) [37], derived from measurements in a broader temperature range, are of similar order as those from the present work (150 and 124 meV, respectively). The gap value of 120 meV, estimated from optical measurements of GaTa$_4$Se$_8$ [40,65], also is compatible with the present result. Assuming the $P\bar{4}2_1m$ space group for the disputed low-temperature phase of GaTa$_4$Se$_8$, Zhang et al. [43] have recently deduced an energy gap of 20 meV from DFT calculations without having to assume any Hubbard $U$. The discrepancy to the present results indicating much higher $E_g$ may point to a different low-temperature structure of GaTa$_4$Se$_8$ (cf. section I).

Antipolar materials should not exhibit any macroscopic spontaneous polarization $P$ because the oppositely oriented dipoles cancel each other out. Only at high electrical dc fields $E$, a $P(E)$ double hysteresis should show up, reflecting the switching into a state with parallel alignment of all dipoles along the field [66]. Such switching is prerequisite for calling a material antiferroelectric. However, often its detection is not possible because the non-zero conductivity of the sample and electrical breakdown prevent the application of the necessary high fields. Just as previously reported for GaNb$_4$S$_8$ [31], we indeed found it impossible to apply sufficiently high fields to GaNb$_4$Se$_8$ and GaTa$_4$Se$_8$ samples to detect the double hysteresis. Polarization measurements were solely possible around and below $T_m \approx 31$ K of the Nb compound as only here the sample conductivity was sufficiently low to allow for such measurements. The results are shown in Fig. 6(a) for various positive and negative fields up to 11.6 kV/cm. Against the expectation for an antipolar material, these data reveal a small, switchable polarization starting to arise at $T < T_m$ and saturating at lower temperatures. This indicates weak ferroelectric dipole ordering. The absolute values of the saturation polarization are much smaller than those reported for polar lacunar spinels [10,11,12,24]. Moreover, part of the detected polarization is generated by thermally stimulated discharge currents (TSDC) arising from charge carriers trapped upon cooling with applied poling fields [67,68]. This is evidenced by cooling-rate-dependent pyrocurrent experiments as shown in Fig. 6(b). Here the broad main peak shifts to higher temperatures for higher rates, indicating its TSDC origin [67,68]. In contrast, the minor sharp peak close to $T_m$ only reveals a marginal shift in opposite direction and thus can be ascribed to dipolar polarization. For a poling field of 11.5 kV/cm, an analysis of this sharp peak alone yields about ten times smaller values of the saturation polarization (~0.005 μC/cm$^2$ for $E$ = 11.5 kV/cm). Interestingly, the results of Fig. 6 qualitatively resemble those for GaNb$_4$S$_8$ [31]. There, the very small ferroelectric polarization, superimposed on antipolar order, was speculated to arise from polar domain walls or tiny canting of the antipolar order [69,70,71]. It should be noted that the too high conductivity made it impossible to measure pyrocurrent across the quadrupolar transition observed at higher temperatures in GaNb$_4$Se$_8$. Thus, on an experimental basis we cannot exclude that weak ferroelectricity may already occur below $T_Q$, though it would not be compatible with the cubic chiral symmetry reported in [31].

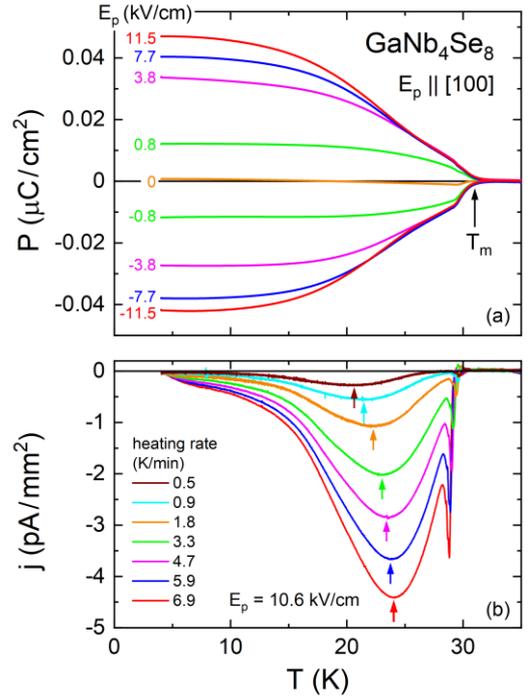

FIG. 6. (a) Temperature-dependent polarization of GaNb$_4$Se$_8$ for various electric poling fields. (b) Temperature dependence of the pyroelectric current $j$ for different heating rates after poling with 10.6 kV/cm. The arrows in (b) indicate the rate-dependent shift of the main hump position.

## IV. SUMMARY AND CONCLUSIONS

In summary, our detailed study of the lacunar spinels GaNb$_4$Se$_8$ and GaTa$_4$Se$_8$ using dielectric-spectroscopy,

polarization, resistivity, magnetic-susceptibility, and specific-heat measurements reveal clear anomalies at the phase transitions of these materials in all measured quantities. In contrast to earlier specific-heat investigations of polycrystalline samples [41,49], our present measurements on single crystals lead to much sharper peaks at these transitions, supporting their first-order nature. Dielectric spectroscopy performed in a broad frequency range enables the unequivocal deconvolution of intrinsic and non-intrinsic dielectric properties [51,52]. Aside from hints at dipolar relaxation dynamics and hopping conductivity, the most interesting finding is the occurrence of downward steps of the intrinsic $\varepsilon'(T)$ when crossing the phase transitions of both materials upon cooling. This indicates antipolar ordering accompanying the structural and magnetic changes at the phase transitions [34,54]. Remarkably, both transitions occurring in $GaNb_4Se_8$ reveal such an anomaly, implying two successive dipolar rearrangements. Hints at antipolar order were previously also found at the phase transition of $GaNb_4S_8$ [31], which is mainly driven by the JT effect, in contrast to the present systems where spin-orbit coupling is believed to play an important role. Due to the significant conductivity of $GaNb_4Se_8$ and $GaTa_4Se_8$, polarization measurements could only be performed around the low-temperature, magneto-structural transition in the Nb compound. Just as for $GaNb_4S_8$, they reveal weak ferroelectricity, likely due to polar domain walls or weak dipole canting.

The dc resistivity of both materials shows thermally activated behavior with significant changes of the corresponding energy gaps and an increase of the resistivity at the transitions. Especially for $GaTa_4Se_8$, this increase is huge (about three decades) although the activation energy becomes reduced at the transition. A likely explanation is a crossover to charge transport dominated by electrons (or holes) excited from in-gap impurity levels to the conduction band instead of the excitation of electrons across a band gap. In this context, it is interesting that the intriguing resistive switching found in several lacunar spinels, including $GaTa_4Se_8$ [39,40] can be well understood when assuming in-gap electronic states arising from impurities or defects [72].

Overall, the findings of the present work show that antipolar ordering in lacunar spinels, triggered by distortions of the cubane units, most likely is not limited to $GaNb_4S_8$ [31] and occurs irrespective of the microscopic origin of the corresponding phase transition. Further work should be especially devoted to the clarification of the charge-transport mechanisms in the low-temperature phases of $GaNb_4Se_8$ and $GaTa_4Se_8$ and to the origin of the resistivity changes at their phase transitions, which may also help to understand the prominent resistive switching in these and related systems.

## ACKNOWLEDGMENTS

This work was supported by the Deutsche Forschungsgemeinschaft through the Transregional Collaborative Research Center TRR 80. The support by the project ANCD 20.80009.5007.19 (Moldova) is also acknowledged.

## APPENDIX A: MAGNETIC SUSCEPTIBILITY AT LOW TEMPERATURES

As mentioned above, the low-temperature upturn of $\chi(T)$ revealed in Figs. 1(b) and 2(b) can be ascribed to small amounts of impurities or defects [41]. To roughly estimate their concentration, we have fitted the experimental data by a Curie law. We found that a pure Curie law at best could describe the experimental data below about 10 K only. Thus, we applied an additional temperature-independent term, $\chi_0$, formally accounting for other contributions to $\chi(T)$. This leads to reasonable fits extending closer to the magnetic transition $T_m$ (lines in Fig. 7). Assuming impurities or defects with $S = 1/2$, from the obtained Curie constants $C = 7.7 \times 10^{-3}$ and $3.3 \times 10^{-3}$ emu K/mol, we arrive at fractions of about 2 and 0.9 % for the Nb and Ta compounds, respectively. This is of similar order of magnitude as found for polycrystalline samples in Ref. [41].

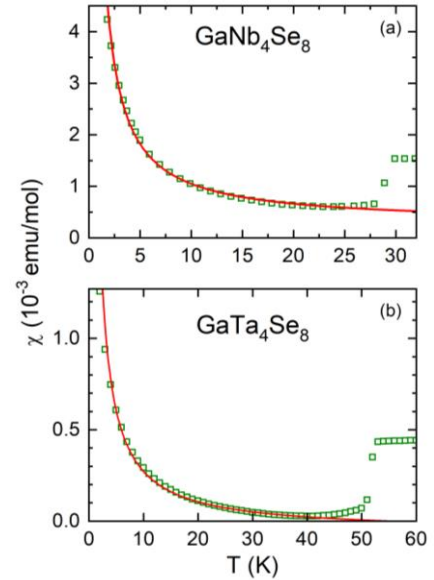

FIG. 7. Temperature dependence of the magnetic susceptibility of $GaNb_4Se_8$ (a) and $GaTa_4Se_8$ (b) at low temperatures. The lines are fits by $C/T + \chi_0$.

One may ask whether these magnetic impurities/defects may also be responsible for the in-gap states invoked above to rationalize the $\rho_{dc}(T)$ behavior at low temperatures. The number of available charge carriers that, in principle, may be excited into the conduction band enters Eq. (1) via the prefactor $\rho_0$, which should be inversely proportional to it. Below $T_m$, $\rho_0$ increases by 10 or by 5 decades for Nb and Ta, respectively (cf. Fig. 5). When assuming that the other quantities contributing to $\rho_0$ (e.g., the charge-carrier mobility) remain unchanged at the transition, this implies that the magnetic defects giving rise to the Curie upturn in $\chi(T)$ are not responsible for the resistivity below $T_\text{m}$. However, one has to admit that various assumptions had to be made to arrive at this conclusion.

---